\documentclass[10pt,twocolumn,twoside,journal]{IEEEtran}

\usepackage[noadjust]{cite}

\ifCLASSINFOpdf
  \usepackage[pdftex]{graphicx}
  \DeclareGraphicsExtensions{.pdf,.jpeg,.png,.tif,.tiff}
\else

\fi

\usepackage[cmex10]{amsmath}

\ifCLASSOPTIONcompsoc
    \usepackage[tight,normalsize,sf,SF]{subfigure}
\else
    \usepackage[tight,footnotesize]{subfigure}
\fi

\usepackage[latin1]{inputenc}
\usepackage{amsthm, amsfonts, amssymb, mathrsfs, amsmath}
\usepackage{amsmath}
  {
      \theoremstyle{plain}
      
  }
\usepackage[pdftex]{graphicx}
\usepackage{nicefrac}
\usepackage{mathtools}
\usepackage{multirow}
\usepackage[ruled]{algorithm2e}
\usepackage[mathcal]{euscript}

\usepackage[draft]{fixme}
\usepackage{enumerate}
\usepackage{dsfont}

\usepackage{tikz}
\allowdisplaybreaks[1]

\newcommand{\leqnomode}{\tagsleft@true}
\newcommand{\reqnomode}{\tagsleft@false}
\makeatother

\newcommand{\bM}{{\boldsymbol{M}}}
\newcommand{\bL}{{\boldsymbol{L}}}
\newcommand{\bS}{{\boldsymbol{S}}}
\newcommand{\bZ}{{\boldsymbol{Z}}}
\newcommand{\bB}{{\boldsymbol{B}}}

\newcommand{\Ojc}{{\Omega_{jc}}}

\newcommand{\bG}{{\boldsymbol{\gamma}}}

\newcommand{\bT}{{\boldsymbol{\Theta}}}

\newcommand{\bU}{{\boldsymbol{U}}}
\newcommand{\bV}{{\boldsymbol{V}}}
\newcommand{\bSi}{{\boldsymbol{\Sigma}}}
\newcommand{\bX}{{\boldsymbol{X}}}
\newcommand{\bx}{{\boldsymbol{x}}}

\newcommand{\bv}{{\boldsymbol{v}}}
\newcommand{\bz}{{\boldsymbol{z}}}
\newcommand{\by}{\boldsymbol{y}}

\newcommand{\bhx}{{\boldsymbol{\widehat{x}}}}
\newcommand{\bhv}{{\boldsymbol{\widehat{v}}}}

\makeatletter
\renewcommand{\@algocf@capt@plain}{above}
\makeatother

\DeclareMathOperator*{\argmin}{arg\,min}


\usepackage{amsthm}

\begin{document}
\title{Online  Decomposition of Compressive Streaming Data Using $n$-$\ell_1$ Cluster-Weighted Minimization}

\author{Huynh Van Luong,~Nikos Deligiannis,~S{\o}ren Forchhammer,~and~Andr\'{e} Kaup

\thanks{H. V. Luong and A. Kaup are with the Chair of Multimedia Communications and Signal Processing, Friedrich-Alexander-Universit\"{a}t Erlangen-N\"{u}rnberg, 91058 Erlangen, Germany (e-mail: huynh.luong@fau.de and andre.kaup@fau.de).}
\thanks{N. Deligiannis is with the Department of Electronics and Informatics, Vrije Universiteit Brussel, 1050 Brussels, and also with iMinds, 9050 Ghent, Belgium (e-mail: ndeligia@etro.vub.ac.be).}
\thanks{S. Forchhammer is with the Department of Photonics Engineering, Technical University of Denmark, 2800 Lyngby, Denmark (e-mail: sofo@fotonik.dtu.dk).}
}

%
\maketitle

\begin{abstract}
	We consider a decomposition method for compressive streaming data in the context of online compressive Robust Principle Component Analysis (RPCA). The proposed decomposition solves an $n$-$\ell_1$ cluster-weighted minimization to decompose a sequence of frames (or vectors), into sparse and low-rank components, from compressive measurements. Our method processes a data vector of the stream per time instance from a small number of measurements in contrast to conventional batch RPCA, which needs to access full data. The $n$-$\ell_1$ cluster-weighted minimization leverages the sparse components along with their correlations with multiple previously-recovered sparse vectors. Moreover, the proposed minimization can exploit the structures of sparse components via clustering and re-weighting iteratively. The method outperforms the existing methods for both numerical data and actual video data. 
\end{abstract}
\vspace{-0.9pt}
\section{Introduction}\label{sec:intro}
\vspace{-0.7pt}

Robust principal component analysis (RPCA)~\cite{CandesRPCA} has been a useful tool for data analysis and inference in many applications, e.g., web data analysis and computer vision \cite{CandesRPCA}. Formally, RPCA models a data matrix~$\bM\in\mathbb{R}^{n\times t}$ as the sum of a \textit{sparse} component matrix~$\bX$ and a \textit{low-rank} component matrix $\bL$ and solves the principal component pursuit~(PCP) \cite{CandesRPCA}~problem:
\vspace{-0.7pt}
\begin{equation}\label{PCP}
\min_{\bL,\bS} \|\bL\|_{*}+\lambda\|\bX\|_{1} \text{~s.t.~}\bM=\bL+\bX,
\vspace{-0.7pt}
\end{equation}
where $\|\bL\|_{*}=\sum_i\sigma_i(\bL)$ is the nuclear norm---sum of singular
values~$\sigma_i(\bL)$---of the matrix~$\bL$, $\|\bX\|_1$ is the $\ell_1$-norm of~$\bX$ (seen as a long vector), and~$\lambda$ is a balance parameter. This batch method requires access to the full data $\bM$, where the low-rank $\bL$ lies on the low-dimensional subspace and the sparse $\bX$ accounts for structured discrepancies. In video analysis, a sequence of vectorized frames (modeled by~$\bM$) is separated into the slowly-changing background~$\bL$ and the sparse foreground $\bX$. However, the assumptions of full data and the static background may be invalid in typical long video streams with background variations.

The online RPCA method \cite{GuoQV14,Rodriguez16} and its compressive counterpart \cite{JWright13,JHe12,pan2017online} have been proposed to process as each column in $\bM$ from compressive measurements. These approaches \cite{GuoQV14,Rodriguez16} assume slow-variation of the low-rank component and leverage compressed sensing (CS)~\cite{CandesTIT06,DonohoTIT06} to recover the sparse component. 
Unlike batch~\cite{CandesRPCA} or online~\cite{Rodriguez16} RPCA approaches, the methods in \cite{JWright13,JHe12} operate on compressive measurements to tackle the computational issues and reduce the cost of data communication and storage. The method in \cite{pan2017online} simultaneously does both on compressive measurements and online. However, these methods do not explore prior kowledge expressing the correlations between the incoming components and prior decomposed vectors.

The problem of reconstructing a series of time-varying sparse signals using prior information has been explored in online RPCA~\cite{GuoQV14} and recursive CS~\cite{MotaTIT17,MotaTSP16}. The study in~\cite{GuoQV14} proposed a recursive method to the compressive case and used modified-CS~\cite{NVaswaniTSP10} to leverage prior support knowledge under the condition of slowly-varying support. 
The study in~\cite{MotaTSP16} assumed the low-rank components non-varying and recovered the sparse component using $\ell_1$-$\ell_1$ minimization ~\cite{MotaTIT17}. However, these methods do not exploit multiple prior information from multiple previously recovered frames.

The use of structural sparse components as prior knowledge has been studied in \cite{VCevher09,RGBaraniuk10,CWLim14,CHegde15}. Model-based CS \cite{RGBaraniuk10} showed that prior information of the signal structure can be used to reduce the number of measurements. The study in \cite{CHegde15} introduced approximation algorithms to extend model-based CS to a wider class of signal models, whereas, the work in \cite{CWLim14} leveraged the support of periodic clustered sparse signals. Alternatively, the structured sparsity model \cite{VCevher09}, which constrains signal coefficients into $C$-clusters without assuming prior knowledge of the locations and sizes of the clusters, has provided provable performance guarantees. Motivated by these ideas, we aim at not only exploiting the clustered-based model for multiple prior information but also re-weighting the clustered sparse components per iteration during the decomposition process.

\textbf{Problem.} We consider a compressive online decomposition method that recursively decomposes streaming data from compressive measurements by leveraging multiple previously decomposed data priors. At time instance $t$, we aim to decompose $\bM_{t}=\bL_{t}+\bX_{t}\in\mathbb{R}^{n\times{t}}$ into $\bX_{t}=[\bx_{1}~ \bx_{1}~...~\bx_{t}]$ and $\bL_{t}=[\bv_{1}~\bv_{2}~...~ \bv_{t}]$, where $\bx_{t}, \bv_{t}\in \mathbb{R}^{n}$ are column-vectors in $\bX_{t}$ and $\bL_{t}$, respectively. We assume that $\bL_{t-1}=[\bv_{1}~\bv_{2}~ ...~\bv_{t-1}]$ and $\bX_{t-1}=[\bx_{1}~\bx_{1}~...~\bx_{t-1}]$ have been recovered at $t-1$ and that at time instance $t$ we have access to compressive measurements $\by_{t}=\mathbf{\Phi}(\bx_{t}+\bv_{t})$, where $\mathbf{\Phi}\in\mathbb{R}^{m\times n} $~$(m\ll n)$ is a random projection~\cite{CandesTIT06}. At time instance $t$, we formulate the decomposition problem 
\begin{align}\label{onlinePCP}
&\underset{\bx_t,\bv_t}{\min}\Big\{ 
\Big\|[\bL_{t-1}~\bv_{t}]\Big\|_{*}+\lambda_1\|\bx_{t}\|_{1} + \lambda_2f_{\mathrm{prior}}(\bx_{t},\bX_{t-1}) \Big\}\nonumber\vspace{-17pt}
\\
\vspace{-15pt}
&~\text{s.t.}~\by_{t}=\mathbf{\Phi}(\bx_{t}+\bv_{t}),
\vspace{-46pt}
\end{align}
where~$f_{\mathrm{prior}}(\cdot)$ expresses the relation of $\bx_t$ with the previously recovered sparse components~$\bX_{t-1}$. In essence, Problem~\eqref{onlinePCP} have been formulated to exploit temporal correlation across multiple priors, e.g., the backgrounds and the foregrounds in multiple video frames. Moreover, we want to leverage the structures in $\bx_t$ and $\bX_{t-1}$ to reduce further the number of measurements.

\textbf{Contribution}. We propose a \textit{compressive online decomposition algorithm} (CODA) that solves Problem~\eqref{onlinePCP} via an $n$-$\ell_1$ cluster-weighted minimization. The algorithm recovers recursively the low-rank and sparse vectors by using the $n$-$\ell_1$ minimization~\cite{LuongGlobalSIP17} given multiple prior information. CODA also leverages the structures of the sparse components by iteratively clustering the sparse components and re-weighting them accordingly in the $n$-$\ell_1$ cluster-weighted minimization. 


\vspace{-0.10pt}
\section{Background}
\label{relatedWork}
\vspace{-0.10pt}
We review fundamental recovery problems \cite{DonohoTIT06,MotaTIT17,LuongICIP16,CandesRPCA} related to our work. Let $\bx\in\mathbb{R}^{n}$ denote a sparse signal for which we have access to random Gaussian measurements~$\by=\mathbf{\Phi}\bx\in\mathbb{R}^{m}$, with~$m\ll~n$. According to the CS theory~\cite{DonohoTIT06}, $\bx$ can be recovered by solving: $\min_{\bx} ||\bx||_{1} \mathrm{~s.t.~} \by\hspace{-2pt}=\hspace{-2pt}\mathbf{\Phi}\bx$
that can be written as 
\vspace{-0.5pt}
\begin{equation}\label{l1-general}
\min_{\bx}\{f(\bx) + g(\bx)\},
\vspace{-0.6pt}
\end{equation}
where $g(\bx)=\lambda ||\bx||_{1}$, $\lambda>0$ is a regularization parameter, and $f(\bx)=\frac{1}{2}||\mathbf{\Phi}\bx-\by||^{2}_{2}$. By using a proximal gradient method \cite{Beck09}, $\bx^{(k)}$ at iteration $k$ can be iteratively computed as
\vspace{-0.6pt}
\begin{equation}\label{l1-proximal}
\bx^{(k)}= \Gamma_{\frac{1}{L}g}\Big(\bx^{(k-1)}\hspace{-2pt}-\hspace{-2pt}\frac{1}{L}\nabla f(\bx^{(k-1)})\Big),
\vspace{-0.6pt}
\end{equation}
where $L\hspace{-2pt}\geq \hspace{-2pt}L_{\nabla f}$\hspace{-1pt} is the Lipschitz constant and $\Gamma_{\frac{1}{L}g}(\bx)$ is a proximal operator defined as
\vspace{-0.4pt}
\begin{equation}\label{l1-proximalOperator}
\Gamma_{\frac{1}{L}g}(\bx) = \argmin_{\bv \in\mathbb{R}^{n}}\Big\{ \frac{1}{L}g(\bv) + \frac{1}{2}||\bv-\bx||^{2}_{2}\Big\}.
\vspace{-0.6pt}
\end{equation}

Alternatively, the $\ell_{1}$-$\ell_{1}$ minimization problem \cite{MotaTIT17} attempts to reconstruct $\bx$ given a side information signal $\bz \in \mathbb{R}^{n}$ by solving Problem \eqref{l1-general} with $g(\bx)=\lambda (\|\bx\|_{1}+\|\bx-\bz\|_{1})$, that is,
\begin{equation}\label{l1-l1minimization}
\min_{\bx}\Big\{\frac{1}{2}\|\mathbf{\Phi}\bx-\by\|^{2}_{2} + \lambda (\|\bx\|_{1}+\|\bx-\bz\|_{1})\Big\}.
\end{equation}

The algorithm in \cite{LuongICIP16} addresses an $n$-$\ell_1$ minimization problem that 
minimize the objective function following in~\eqref{l1-general} by:
\vspace{-0.8pt}
\begin{equation}\label{n-l1minimizationGlobal}
\min_{\bx}\hspace{-2pt}\Big\{\frac{1}{2}||\mathbf{\Phi}\bx-\by||^{2}_{2} + \lambda \hspace{-2pt}\sum\limits_{j=0}^{J}\hspace{-2pt}\beta_{j}||\mathbf{W}_{j}(\bx-\bz_{j})||_{1}\Big\}.
\vspace{-0.9pt}
\end{equation}
where $\bx$ is the signal to be recovered, $\bz_0=\mathbf{0}$ and $\bz_1,\dots,\bz_J$ are $J$ prior information signals, $\beta_{j}\hspace{-2pt}>\hspace{-2pt}0$ are weights across the prior information vectors, and $\mathbf{W}_{j}=\mathrm{diag}(w_{j1},w_{j2},...,w_{jn})$, with $w_{ji}>0$, is a diagonal matrix weighting each element $i\in\{1,\dots,n\}$ of each prior information vector $\bz_j$. \textcolor{black}{It is worth noting that $\bz_0=\mathbf{0}$ is to promote the sparsity of $\bx$.}


The PCP \cite{CandesRPCA} problem subsumes the CS\ problem. To show this, we follow the formulation in~\eqref{l1-general} and write Problem \eqref{PCP} as
\vspace{-0.6pt}
\begin{equation}\label{generalPCP}
\min_{\bL,\bS}\{\mathcal{F}(\bL,\bX) + \mathcal{G}(\bL,\bX)\},
\vspace{-0.10pt}
\end{equation}
where $\mathit{\mathcal{F}}(\bL,\bX)=\frac{1}{2}\|\bM-\bL-\bX\|^{2}_{F}$ and $\mathcal{G}(\bL,\bX)=\mu\|\bL\|_{*}+\mu\lambda\|\bX\|_{1}$, with $\|\cdot\|_{F}$ denoting the Frobenious norm. Using proximal gradient methods, \cite{Beck09} gives that $\bL^{(k+1)}$ and $\bX^{(k+1)}$ at iteration $k+1$ can be iteratively computed via the singular value thresholding operator \cite{Cai10} for $\bL$ and the soft thresholding operator \cite{Beck09} for $\bX$.

\vspace{-0.12pt}
\section{Compressive Online Decomposition Using $n$-$\ell_1$ Cluster-Weighted Minimization}
\vspace{-0.8pt}
\label{RAMSIA}

\subsection{The $n$-$\ell_1$ Cluster-Weighted Minimization Problem}\label{problem}
\vspace{-0.6pt}
%
%
%
%
\vspace{-0.2pt}
The proposed $n$-$\ell_{1}$ cluster-weighted minimization is based on our previous work \cite{LuongGlobalSIP17} and enhanced by promoting the natural structures of data. At time instance $t$, the method receives as input compressive measurements $\by_{t}=\mathbf{\Phi}(\bx_{t}+\bv_{t})\in\mathbb{R}^{m}$ of a data vector and estimates the sparse and low-rank components ($\bhx_{t}$, $\bhv_{t}$, respectively) with the aid of prior information. All elements of each vector $\bx_t-\bz_j$ are clustered into $C$ subsets. Let $\Ojc \subseteq  \{1,~ 2,...,~n\}$ denote a set of indices of elements belonging to cluster $c\in\{1,...,C\}$, that constitute a cluster vector $\bx_{t|_{\Omega_{jc}}}-\bz_{j|_{\Omega_{jc}}}$, and $n_{jc}$ denote the number of elements of the cluster $\Ojc$, denoted as $n_{jc}=|\Omega_{jc}|$, i.e., $\bx_{t|_{\Omega_{jc}}}-\bz_{j|_{\Omega_{jc}}}\in \mathbb{R}^{n_{jc}}$. In this way, $\sum_{c=1}^{C}n_{jc}=n$. We denote $\boldsymbol{\gamma}_{j}=\mathrm{diag}(\gamma_{j1},\gamma_{j2},...,\gamma_{jn})$, with $\gamma_{ji}\in \mathbb{R}^{+}$, a diagonal matrix weighting each element $i\in\{1,\dots,n\}$ of each prior information vector $\bz_{j}$. We assign that for each cluster $\Omega_{jc}$ all components of $\boldsymbol{\gamma}_{j|_{\Omega_{jc}}}$ are equal to $\bar{\gamma}_{jc} \in \mathbb{R}^{+}$, i.e., for any index belonging to cluster $c$, $i\in \Omega_{jc}$, $\gamma_{ji|_{\Omega_{jc}}}=\bar{\gamma}_{jc}$. The method solves the following problem:
\vspace{-0.10pt}
\begin{align}\label{CORPCAminimization}
\hspace{-4pt}\min_{\bx_{t},\bv_{t}}\hspace{-2pt}\Big\{&\hspace{-2pt}H(\bx_{t},\bv_{t})\hspace{-2pt}
=\hspace{-2pt}\frac{1}{2}\|\mathbf{\Phi}(\bx_{t}\hspace{-2pt}\nonumber\\
&+\hspace{-2pt}\bv_{t})\hspace{-2pt}-\hspace{-2pt}\by_{t}\|^{2}_{2}\hspace{-2pt}+\hspace{-2pt}\lambda \mu\hspace{-2pt}\sum\limits_{j=0}^{J}\hspace{-2pt}\beta_{j}\|\boldsymbol{\gamma}_{j}\mathbf{W}_{j}(\bx_{t}\hspace{-2pt}-\hspace{-2pt}\bz_{j})\|_{1}\hspace{-2pt}\nonumber\\
&+\hspace{-2pt}\mu\Big\|[\bB_{t-1}~ \bv_{t}]\Big\|_{*}\hspace{-2pt}\Big\},
\vspace{-0.8pt}
\vspace{-0.19pt}
\vspace{-0.45pt}
\end{align}
Decomposing further the above problem into a cluster-based formulation as
\vspace{-0.5pt}
\begin{align}\label{CORPCAminimizationC}
\hspace{-2pt}\min_{\bx_{t},\bv_{t}}\hspace{-2pt}\Big\{\hspace{-2pt}H(\bx_{t},\bv_{t})\hspace{-2pt} 
&=\hspace{-0pt}\frac{1}{2}\|\mathbf{\Phi}(\bx_{t}+\bv_{t})-\by_{t}\|^{2}_{2}\nonumber \\
&+\lambda \mu\hspace{-2pt}\sum\limits_{j=0}^{J}\hspace{-2pt}\beta_{j}\hspace{-2pt}\sum\limits_{c=1}^{C}\hspace{-2pt}\bar{\gamma}_{jc}\|\mathbf{W}_{j|_{\Omega_{jc}}}\hspace{-2pt}(\bx_{t|_{\Omega_{jc}}}\hspace{-2pt}-\hspace{-2pt}\bz_{j|_{\Omega_{jc}}})\|_{1}\nonumber\\
&+\mu\Big\|[\bB_{t-1}~ \bv_{t}]\Big\|_{*}\Big\},
\vspace{-0.8pt}
\vspace{-0.19pt}
\vspace{-0.40pt}
\end{align}
where $\lambda$, $\mu>0$ are tuning parameters, and $\bZ_{t-1}:=\{\bz_{j}\}_{j=1}^J$, $\bB_{t-1}\in
\mathbb{R}^{n\times d}$ are matrices that serve as prior
information for $\bx_{t}$ and $\bv_{t}$, respectively. The components in~$\bZ_{t-1}$ and $\bB_{t-1}$ can be
a direct (sub-)set of the previously reconstructed data vectors $\{ \hat\bx_{1},
..., \hat\bx_{t-1}\} $ and $\{\hat\bv_{1}, ..., \hat\bv_{t-1}\}$, or formed after applying a processing step. In the case of video data, the processing
step can compensate for the motion across the frames \cite{LuongSoICT17} by means
of optical flow~\cite{TBrox11}. 
\setlength{\textfloatsep}{0pt}
\begin{algorithm}[t!] 
	\DontPrintSemicolon \SetAlgoLined
	\textbf{Input}: $\by_{t},~\bZ_{t-1},~\bB_{t-1}$;\\
	\textbf{Output}: $\bhx_{t},~\bhv_{t},~\bZ_{t},~\bB_{t}$;\\
	\tcp{Initialize variables and parameters.}
	$\bx_{t}^{(-1)}\hspace{-2pt}=\hspace{-2pt}\bx_{t}^{(0)}\hspace{-2pt}=\hspace{-2pt}\mathbf{0}$; $\bv_{t}^{(-1)}\hspace{-2pt}=\hspace{-2pt}\bv_{t}^{(0)}\hspace{-2pt}=\hspace{-2pt}\mathbf{0}$;
	$\xi_{-1}\hspace{-2pt}=\xi_{0}\hspace{-2pt}=\hspace{-2pt}1$; $\mu_{0}\hspace{-2pt}=\hspace{-2pt}0$; $\bar{\mu}\hspace{-2pt}>\hspace{-2pt}0$; $\lambda>0$; $0\hspace{-2pt}<\hspace{-2pt}\epsilon\hspace{-2pt}<\hspace{-2pt}1$; $k\hspace{-2pt}=\hspace{-2pt}0$; $g_{1}(\cdot)\hspace{-2pt}=\hspace{-2pt}\|\cdot\|_{1}$; $\mathbf{\Phi}$;\\
	\While{not converged}{\label{converged}
		\tcp{Solve Problem \eqref{CORPCAminimization}.}
		$\widetilde{\bv_{t}}^{(k)}\hspace{-2pt}=\bv_{t}^{(k)}\hspace{-2pt}+\hspace{-2pt}\frac{\xi_{k-1}-1}{\xi_{k}}(\bv_{t}^{(k)}\hspace{-2pt}-\hspace{-2pt}\bv_{t}^{(k-1)})$; \label{startV}\\
		$\widetilde{\bx_{t}}^{(k)}\hspace{-2pt}=\bx_{t}^{(k)}\hspace{-2pt}+\hspace{-2pt}\frac{\xi_{k-1}-1}{\xi_{k}}(\bx_{t}^{(k)}\hspace{-2pt}-\hspace{-2pt}\bx_{t}^{(k-1)})$; \\
		$\nabla_{\bv_{t}} f(\widetilde{\bv_{t}}^{(k)},\widetilde{\bx_{t}}^{(k)})=\nabla_{\bx_{t}}f(\widetilde{\bv_{t}}^{(k)},\widetilde{\bx_{t}}^{(k)})=\mathbf{\Phi}^{\mathrm{T}}\Big(\mathbf{\Phi} (\widetilde{\bv_{t}}^{(k)}+\widetilde{\bx_{t}}^{(k)})-\by_{t}\Big)$; \\ 
		
		%
		%

		\hspace{-0pt}$(\bU_{t},\bSi_{t},\bV_{t})\hspace{-1pt}=\hspace{-1pt}\hspace{-0pt}\mathrm{incSVD}\Big(\hspace{-0pt}\Big[\bB_{t-1}~\Big(\widetilde{\bv_{t}}^{(k)}\hspace{-2pt}-\hspace{-2pt}\frac{1}{2}\nabla_{\bv_{t}} f(\widetilde{\bv_{t}}^{(k)},\widetilde{\bx_{t}}^{(k)})\Big)\Big]\Big)$; \textcolor{black}{where $\mathrm{incSVD}(\cdot)$ is an incremental singular vector decomposition \cite{Brand02}}; \label{incSVD}\\
		
		$\bT_{t}\hspace{-2pt}=\hspace{-2pt}\bU_{t}\boldsymbol{\mathit{\Gamma}}_{\frac{\mu_{k}}{2}\boldsymbol{g}_{1}}(\bSi_{t})\bV_{t}^{T}$; where $\boldsymbol{\mathit{\Gamma}}_{\frac{\mu_{k}}{2}\boldsymbol{g}_{1}}(\cdot)$ is given by \eqref{l1-proximalOperatorMatrix};\label{gamma}\\
		$\bv_{t}^{(k+1)}\hspace{-2pt}=\bT_{t}(:,\mathrm{end})$; \\
		$\bx_{t}^{(k+1)}\hspace{-2pt}=\hspace{-2pt}\Gamma_{\frac{\mu_{k}}{2}g}\Big(\widetilde{\bx_{t}}^{(k)}-\frac{1}{2}\nabla_{\bx_{t}} f(\widetilde{\bv_{t}}^{(k)},\widetilde{\bx_{t}}^{(k)})\Big)$; where $\Gamma_{\frac{\mu_{k}}{2}g}(\cdot)$ is given by \eqref{n-l1-proximalOperatorElementFinalInter};\label{endX}\\
		\tcp{Determine clusters.}
		$\{\Omega_{jc}\}_{c=1}^C=f_{\mathrm{clust}}(\bx^{(k)}_t-\bz_{j},C)$;\label{clustering} $n_{jc}=|\Omega_{jc}|$\label{numCluster}\\		
		\tcp{Compute the updated weights.}
		\vspace{-0.2pt}
		$w_{ji|_{\Omega_{jc}}} =\frac{n_{jc}(|x^{(k+1)}_{ti}-z_{ji}|+\epsilon)^{-1}}{\sum_{l\in{ \Omega_{jc}}}(|x^{(k+1)}_{tl}-z_{jl}|+\epsilon)^{-1}}$;\label{weightW}
		\vspace{-0.1pt}
		\\
		$\bar{\gamma}_{jc} =\dfrac{\Big(\|\mathbf{W}_{j|_{\Omega_{jc}}}(\bx^{(k+1)}_{t|_{\Omega_{jc}}}-\bz_{j|_{\Omega_{jc}}})\|_{1}+\epsilon\Big)^{-1}}{\sum\limits_{l=1}^{C}\Big(\|\mathbf{W}_{j|_{\Omega_{jl}}}(\bx^{(k+1)}_{t|_{\Omega_{jl}}}-\bz_{j|_{\Omega_{jl}}})\|_{1}+\epsilon\Big)^{-1}}$;\label{weightGamma}\\
		$\beta_{j} =\dfrac{\Big(\|\boldsymbol{\gamma}_j \mathbf{W}_{j}(\bx^{(k+1)}_t-\bz_{j})\|_{1}+\epsilon\Big)^{-1}}{\sum\limits_{l=0}^{J}\Big(\|\boldsymbol{\gamma}_l \mathbf{W}_{l}(\bx^{(k+1)}_t-\bz_{l})\|_{1}+\epsilon\Big)^{-1}}$;\label{weightBeta}
		
		$\xi_{k+1}=(1+\sqrt{1+4\xi_{k}^{2}})/2$; $\mu_{k+1}=\max(\epsilon\mu_{k},\bar{\mu})$;\\
		$k=k+1$; \\
	}
	\tcp{Update prior information.}
	$\bZ_{t}:=\{\bz_{j}=\bx^{(k+1)}_{t-J+j}\}_{j=1}^{J}$;\label{updateZ}\\
	$\bB_{t}=\bU_{t}(:,1:d)\mathbf{\Gamma}_{\frac{\mu_{k}}{2}g_{1}}(\bSi_{t})(1:d,1:d)\bV_{t}^{\mathrm{T}}(:,1:d)$; \label{updateB}\\
	
	\Return $\bhx_{t}=\bx_{t}^{(k+1)},~\bhv_{t}=\bv_{t}^{(k+1)},~\bZ_{t},~\bB_{t}$;
	\caption{The proposed CODA.}
	\label{CODAAlg}
\end{algorithm}
\vspace{-0.10pt}

\subsection{The Proposed Compressive Online Decomposition Algorithm (CODA)}
\vspace{-0.6pt}
\textbf{Solving Problem \eqref{CORPCAminimizationC}}. CODA solves the $n$-$\ell_{1}$ cluster-weighted minimization problem in \eqref{CORPCAminimizationC} by using proximal gradient methods \cite{Beck09}, where, at every iteration $k$, the algorithm updates the weights $\mathbf{W}_{j}$, $\bG_j$, and $\beta_{j}$, and computes~$\bx$. 
In this
way, we adaptively weight multiple prior information
according to their qualities during the iterative process. In this work,
we set the constraints as $\sum_{i\in \Omega_{jc}}w_{ji}\hspace{-2pt}=\hspace{-2pt}n_{jc}$ for each cluster $c$, $\sum_{c=1}^{C}\bar{\gamma}_{jc}\hspace{-2pt}=\hspace{-2pt}1$ in a given $\bz_j$, and $\sum_{j=0}^{J}\beta_{j}\hspace{-2pt}=\hspace{-2pt}1$ across multiple prior information. Let ~$f(\bv_{t},\bx_{t})=(1/2)\|\mathbf{\Phi}(\bx_{t}+\bv_{t})-\by_{t} \|^{2}_{2}$ and $g(\bx_{t})=\lambda\sum_{j=0}^{J}\beta_{j}\|\bG_j\mathbf{W}_{j}(\bx_{t}-\bz_{j})\|_{1}$.

The algorithm computes $\bx_{t}^{(k+1)}$ and $\bv_{t}^{(k+1)}$ at iteration $k+1$ via the soft thresholding operator \cite{Beck09} and the single value thresholding operator \cite{Cai10}, respectively.
The proximal operator $\boldsymbol{\mathit{\Gamma}}_{\tau \boldsymbol{g}_{1}}(\cdot)$ in Line \ref{gamma} of Algorithm \ref{CODAAlg} is defined as
\vspace{-0.6pt}
\begin{equation}\label{l1-proximalOperatorMatrix}
\boldsymbol{\mathit{\Gamma}}_{\tau \boldsymbol{g}_{1}}(\bX) = \argmin_{\bV }\Big\{ \tau \boldsymbol{g}_{1}(\bV) + \frac{1}{2}||\bV-\bX||^{2}_{F}\Big\},
\vspace{-0.6pt}
\end{equation}
where $\boldsymbol{g}_{1}(\cdot)\hspace{-2pt}=\hspace{-2pt}\|\cdot\|_{1}$.
As keeping $\mathbf{W}_{j}$, $\bG_j$, and $\beta_{j}$ fixed, adhering to the proximal gradient method~\cite{Beck09}, $\bx^{(k+1)}$ is obtained from \eqref{l1-proximal}. 
The proximal operator $\Gamma_{\frac{1}{L}g}(\bx)$ \eqref{l1-proximalOperator} for our problem is given by [we derive as in Appendix in \cite{LuongICIP16}]
\vspace{-0.10pt}
\begin{equation}\label{n-l1-proximalOperatorElementFinalInter}
\Gamma_{\frac{1}{L}g}(x_{ti})=\left\{
\begin{array}{l}
\hspace{-0pt}x_{ti}-\frac{\lambda}{L} \sum\limits_{j=0}^{J}\beta_{j}\gamma_{ji}w_{ji}(-1)^{\mathfrak{b}(l<j)}\mathrm{~~~if\hspace{1pt}}\eqref{n-l1-proximalXAlg}
, \\
z_{li}~~~~~~~~~~~~~~~~~~~~~~~~~~~~~~~~~~~~~\mathrm{if\hspace{1pt}} \eqref{n-l1-proximalZAlg},
\end{array}
\right.
\vspace{-0.8pt}
\end{equation}
with
\vspace{-0.10pt}
\begin{subequations}\label{n-l1-proximalXZAlg}
	\begin{align}
	\hspace{-5pt}z_{li}\hspace{-2pt}+\hspace{-2pt}\frac{\lambda}{L}\hspace{-2pt} \sum\limits_{j=0}^{J}\hspace{-2pt}\beta_{j}\gamma_{ji}w_{ji}&(\hspace{-2pt}-1\hspace{-2pt})^{\mathfrak{b}(l<j)}\hspace{-4pt}<\hspace{-3pt}x_{ti}\hspace{-2pt}<\hspace{-3pt}z_{(l+1)i}\hspace{-2pt}\nonumber\\
	&+\hspace{-2pt}\frac{\lambda}{L}\hspace{-2pt} \sum\limits_{j=0}^{J}\hspace{-2pt}\beta_{j}\gamma_{ji}w_{ji}(\hspace{-2pt}-1\hspace{-2pt})^{\mathfrak{b}(l<j)},
	\vspace{-0.5pt} \label{n-l1-proximalXAlg}\\
	\hspace{-5pt}	z_{li}\hspace{-2pt}+\hspace{-1pt}\frac{\lambda}{L} \hspace{-3pt}\sum\limits_{j=0}^{J}\hspace{-2pt}\beta_{j}\gamma_{ji}w_{ji}&(\hspace{-2pt}-1\hspace{-2pt})^{\mathfrak{b}(l-1<j)}\hspace{-3pt}\leq \hspace{-2pt}x_{ti}\hspace{-3pt}\leq \hspace{-2pt} z_{li}\hspace{-2pt}\nonumber\\
	&+\hspace{-1pt}\frac{\lambda}{L} \hspace{-3pt}\sum\limits_{j=0}^{J}\hspace{-2pt}\beta_{j}\gamma_{ji}w_{ji}(\hspace{-2pt}-1\hspace{-2pt})^{\mathfrak{b}(l<j)}\label{n-l1-proximalZAlg},
	\vspace{-0.13pt}
	\end{align}
	\vspace{-0.3pt}
\end{subequations}
where, without loss of generality, we have assumed that $-\infty =z_{(-1)i}\leq z_{0i}\leq z_{1i}\leq\dots\hspace{-2pt}\leq z_{Ji}\leq z_{(J+1)i}=\infty$, and we have defined a boolean function
\vspace{-0.6pt}
\begin{equation}\label{boolFunction}
\mathfrak{b}(l<j)=
\left\{
\begin{array}{l}
1,\quad \mathrm{if} \quad l<j \\
0, \quad\mathrm{otherwise}.
\end{array}
\right.
\vspace{-0.10pt}
\end{equation}
with $l\in\{-1,\dots, J\}$. It is worth noting that \eqref{n-l1-proximalXAlg} and \eqref{n-l1-proximalZAlg} are disjoint. 

\textbf{Updating weights $\mathbf{W}_{j}$, $\boldsymbol{\gamma}_j$, and $\beta_{j}$}. Firstly, given $\bx_t$, $\beta_{j}$, and $\boldsymbol{\gamma}_j$ (via determining clusters as in Line \ref{clustering} in Algorithm \ref{CODAAlg}, here $f_{\mathrm{cluster}}(\cdot)$ is the k-means clustering algorithm \cite{SLloyd82}), we compute $\mathbf{W}_{j|_{\Omega_{jc}}}$
per prior information $\bz_{j}$ as
\vspace{-0.15pt}
\begin{align}
\label{n-l1-weight-minimizationIntra}
&\arg\min_{\mathbf{W}_{j|_{\Omega_{jc}}}}\{H(\bx_t,\bv_t)\}\nonumber\\
&=\arg\min_{\mathbf{W}_{j|_{\Omega_{jc}}}}\Big\{ \lambda \mu\hspace{-2pt}\sum\limits_{j=0}^{J}\hspace{-2pt}\beta_{j}\sum\limits_{c=1}^{C}\bar{\gamma}_{jc}\|\mathbf{W}_{j|_{\Omega_{jc}}}(\bx_{t|_{\Omega_{jc}}}-\bz_{j|_{\Omega_{jc}}})\|_{1}\Big\}\nonumber\\
&=\arg\min_{\{w_{ji|_{\Omega_{jc}}}\}} \Big\{\sum\limits_{i\in {\Omega_{jc}}}w_{ji}|x_{ti}
-z_{ji}| \Big\},
\vspace{-0.25pt}
\end{align}
where $z_{ji}$ is the $i$-th element of $\bz_j$. Following the Cauchy inequality,
we can minimize~\eqref{n-l1-weight-minimizationIntra} when, for all $i\in \Omega_{jc}$,
$w_{ji}|x_{ti}-z_{ji}|$ is equal to a positive parameter $\eta_{jc}$, i.e., $w_{ji}=\eta_{jc}/(|x_{ti}-z_{ji}|\hspace{-2pt}+\hspace{-2pt}\epsilon)$,
with $\epsilon$ small, \textcolor{black}{such that the zero-valued $|x_{ti}-z_{ji}|$ do not prohibit the iterative computation}. Setting the constraint $\sum_{i\in \Omega_{jc}}w_{ji}\hspace{-2pt}=\hspace{-2pt}n_{jc}$, we
get
\vspace{-0.8pt}
\begin{equation}\label{n-l1-weightsIntra}
w_{ji|_{\Omega_{jc}}} =\frac{n_{jc}(|x_{ti}-z_{ji}|+\epsilon)^{-1}}{\sum\limits_{l\in{ \Omega_{jc}}}(|x_{tl}-z_{jl}|+\epsilon)^{-1}}.
\vspace{-0.8pt}
\end{equation}

Secondly, keeping $\bx_t$, $\beta_{j}$, and $\mathbf{W}_{j|_{ \Omega_{jc}}}$ fixed, we compute $\boldsymbol{\gamma}_j$ via $\bar{\gamma}_{jc}$
as
\vspace{-0.8pt}
\begin{align}\label{n-l1-cluster-weight-minimizationInter}
&\arg\hspace{-0pt}\min_{\{\bar{\gamma}_{jc}\}}\{H(\bx_t,\bv_t)\}\nonumber\\
&=\arg\hspace{-0pt}\min_{\{\bar{\gamma}_{jc}\}}\Big\{\sum\limits_{c=1}^{C}\bar{\gamma}_{jc}\|\mathbf{W}_{j|_{\Omega_{jc}}}(\bx_{t|_{\Omega_{jc}}}-\bz_{j|_{\Omega_{jc}}})\|_{1}\Big\}.
\vspace{-0.8pt}
\end{align}
Similar to~\eqref{n-l1-weight-minimizationIntra}, from \eqref{n-l1-cluster-weight-minimizationInter}
we obtain $\bar{\gamma}_{jc}$ with $\eta_{j}>0$ as
\vspace{-0.8pt}
\begin{equation}\label{n-l1-Gamma}
\bar{\gamma}_{jc}= \eta_{j} \Big/\Big({\|\mathbf{W}_{j|_{\Omega_{jc}}}(\bx_{t|_{\Omega_{jc}}}-\bz_{j|_{\Omega_{jc}}})\|_{1}+\epsilon}\Big).
\vspace{-0.8pt}
\end{equation}
Combining \eqref{n-l1-Gamma} with the constraint $\sum_{c=1}^{C}\bar{\gamma}_{jc}\hspace{-2pt}=\hspace{-2pt}1$,
we get
\vspace{-0.8pt}
\begin{equation}\label{n-l1-BetaInter}
\bar{\gamma}_{jc} =\dfrac{\Big(\|\mathbf{W}_{j|_{\Omega_{jc}}}(\bx_{t|_{\Omega_{jc}}}-\bz_{j|_{\Omega_{jc}}})\|_{1}+\epsilon\Big)^{-1}}{\sum\limits_{l=1}^{C}\Big(\|\mathbf{W}_{j|_{\Omega_{jl}}}(\bx_{t|_{\Omega_{jl}}}-\bz_{j|_{\Omega_{jl}}})\|_{1}+\epsilon\Big)^{-1}}.
\vspace{-0.8pt}
\end{equation}

Finally, keeping $\bx_t$, $\boldsymbol{\gamma}_j$, and $\mathbf{W}_{j}$ fixed, we compute $\beta_{j}$
as
\vspace{-0.12pt}
\begin{equation}\label{n-l1-weight-minimizationInter}
\arg\hspace{-0pt}\min_{\{\beta_{j}\}}\{H(\bx_t,\bv_t)\}= \arg\hspace{-0pt}\min_{\{\beta_{j}\}}\Big\{\lambda
\sum\limits_{j=0}^{J}\beta_{j}\|\boldsymbol{\gamma}_j\mathbf{W}_{j}(\bx_t-\bz_{j})\|_{1}\Big\}.
\vspace{-0.12pt}
\end{equation}
Similar to~\eqref{n-l1-weight-minimizationIntra}, from \eqref{n-l1-weight-minimizationInter}
we obtain $\beta_{j}$ with $\eta_{\beta}>0$ as
$\beta_{j}= \eta_{\beta} /({\|\boldsymbol{\gamma}_j \mathbf{W}_{j}(\bx_t-\bz_{j})\|_{1}+\epsilon}).$
Combining with the constraint $\sum_{j=0}^{J}\beta_{j}=1$,
we get
\vspace{-0.10pt}
\begin{equation}\label{n-l1-BetaInter}
\beta_{j} =\dfrac{\Big(\|\boldsymbol{\gamma}_j \mathbf{W}_{j}(\bx_t-\bz_{j})\|_{1}+\epsilon\Big)^{-1}}{\sum\limits_{l=0}^{J}\Big(\|\boldsymbol{\gamma}_l \mathbf{W}_{l}(\bx_t-\bz_{l})\|_{1}+\epsilon\Big)^{-1}}.
\vspace{-0.10pt}
\end{equation}


CODA [see Algorithm \ref{CODAAlg}] is based on our previous CORPCA\footnote{The code for CORPCA is available at https://github.com/huynhlvd/corpca.} \cite{LuongGlobalSIP17} and operates in two steps: It first solves Problem~\eqref{CORPCAminimization} given $\bZ_{t-1}$ and $\bB_{t-1}\hspace{-1pt}\textcolor{black}{\in\hspace{-1pt}
	\mathbb{R}^{n\times d}}$ and  the reconstructed vectors~$\bhx_{t}$ and $_{t}$ are used to updates $\bZ_{t}$ and $\bB_{t}$, which are to be used in the following time instance. These updates are specified more details in \cite{LuongGlobalSIP17}.

\vspace{-0.6pt}
\section{Experimental Results}\label{Experiments}
\vspace{-0.6pt}
\vspace{-0.2pt}
\begin{figure*}[t!]
	\centering
	\hspace{-45pt}\subfigure[\vspace{-0.45pt}CODA-$n$-$\ell_{1}$]{
		\includegraphics[width=0.58\textwidth]{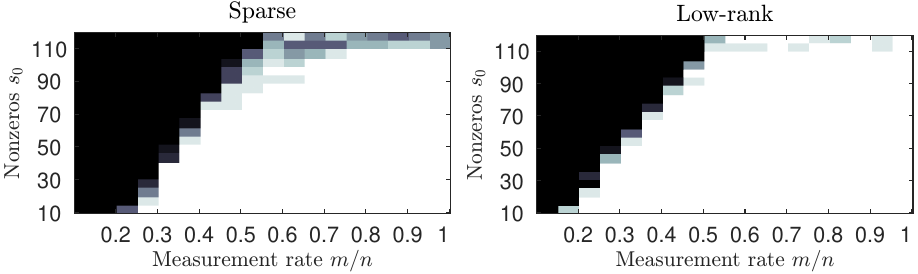}\label{codaPerformBg} } 	
	\hspace{-14pt}	\subfigure[\vspace{-0.15pt}ReProCS \cite{GuoQV14}]{
		\includegraphics[width=0.29\textwidth]{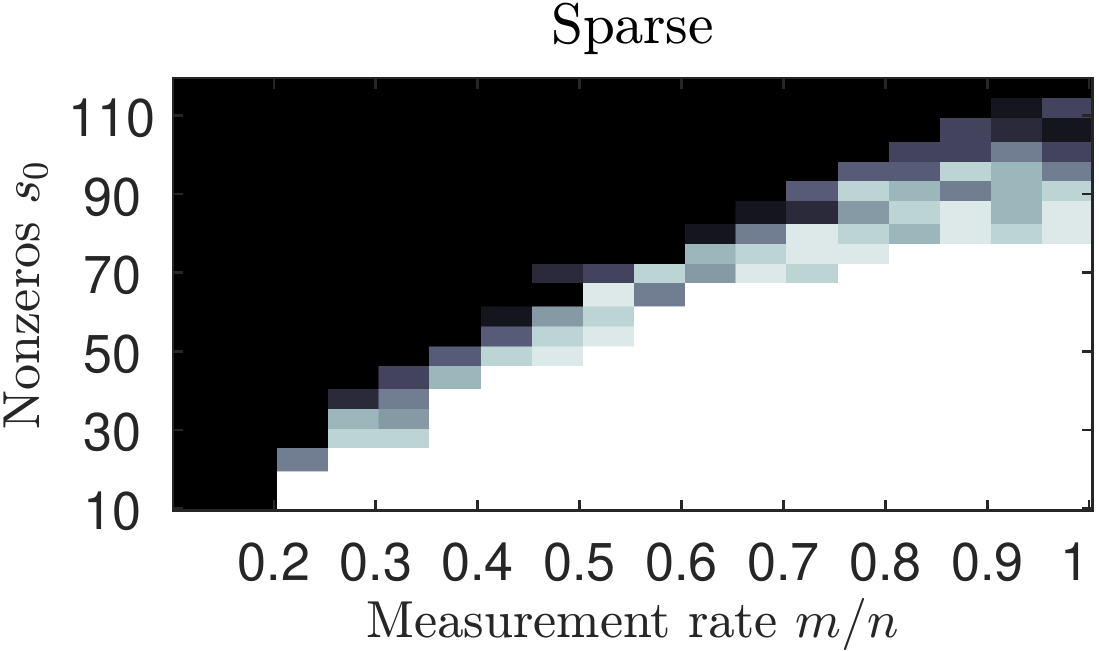}\label{reprocsPerform}} 
	\hspace{-10pt}\subfigure[\vspace{-0.15pt}CORPCA-$n$-$\ell_{1}$ \cite{LuongGlobalSIP17}]{
		\vspace{-0.15pt}
		\includegraphics[width=0.58\textwidth]{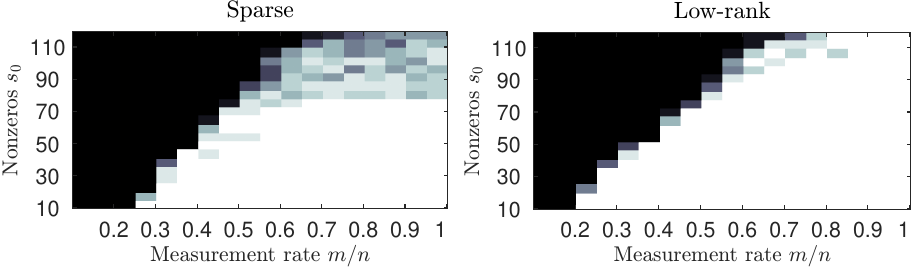}\label{corpcaPerformBg} } 
	\hspace{-8pt}\subfigure[\vspace{-0.15pt}GRASTA \cite{JHe12}]{
		\includegraphics[width=0.29\textwidth]{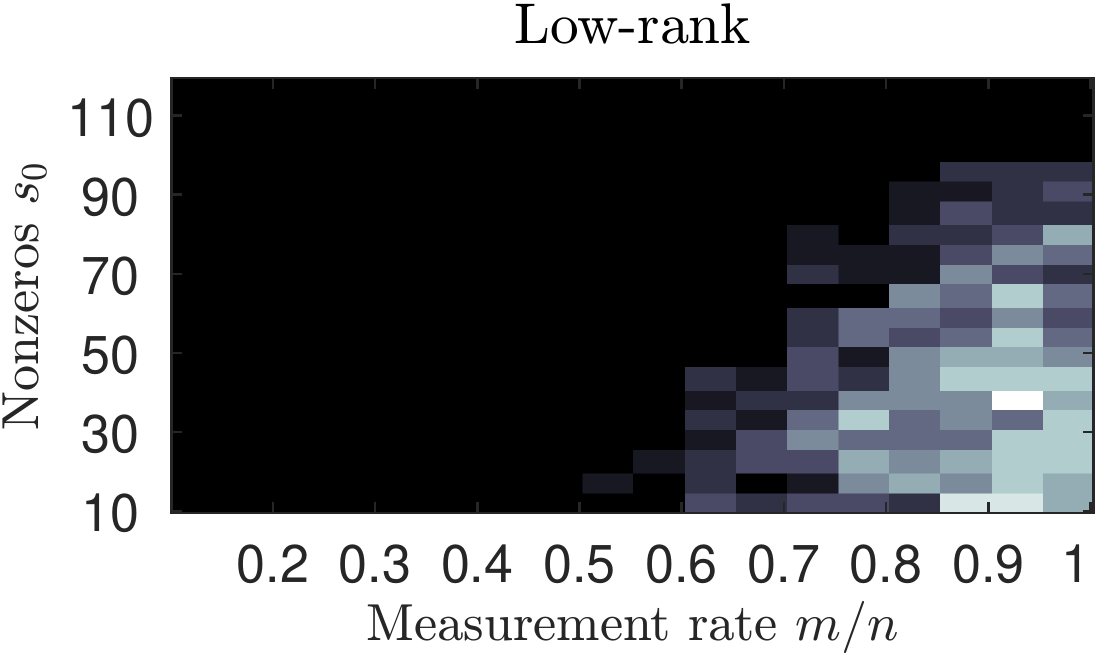}\label{grastaPerform}} 
	\hspace{-1pt}\subfigure[\hspace{-3pt}Scale]{
		\includegraphics[width=0.09\textwidth]{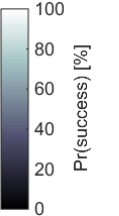}\label{scale}}
	\caption{Average success probabilities for CORPCA, ReProCS~\cite{GuoQV14}, and GRASTA~\cite{JHe12}.
		\vspace{-0.0pt}
	}\label{corpcaPerform}
	\vspace{-0.5pt}
\end{figure*}
\subsection{Performance Using Synthetic Data}\label{corpcaExperiment}
\vspace{-0.0pt}
The performance of Algorithm~\ref{CODAAlg} employing the proposed~$n$-$\ell_1$ cluster-weighted minimization is evaluated and compared to our previous algorithm \cite{LuongGlobalSIP17} with $n$-$\ell_1$ minimization as well as GRASTA \cite{JHe12} and ReProCS \cite{GuoQV14}. GRASTA \cite{JHe12} and ReProCS~\cite{GuoQV14} are online methods, where \textcolor{black}{ReProCS \cite{GuoQV14} recovers the sparse components, while GRASTA recovers the low-rank components \cite{JHe12}.}

We generate our data as follows. First, we generate the low-rank component as $\bL\hspace{-2pt}=\hspace{-2pt}\bU\bV^{\mathrm{T}}$, where $\bU\hspace{-2pt}\in\hspace{-2pt}\mathbb{R}^{n\times r}$ and $\bV\hspace{-2pt}\in\hspace{-2pt}\mathbb{R}^{(d+q)\times r}$ are random matrices whose entries are drawn from the standard normal distribution. We set $n=500$, $r\hspace{-0pt}=\hspace{-0pt}5$, and $d\hspace{-0pt}=\hspace{-0pt}100$ the number of vectors for training and $q\hspace{-0pt}=\hspace{-0pt}100$ the number of testing vectors; this yields $\bL\hspace{-0pt}=\hspace{-0pt}[\bv_{1}\dots\bv_{d+q}]$. Secondly, we generate $\bX\hspace{-0pt}=\hspace{-0pt}[\bx_{1}\dots\bx_{d+q}]$. Specifically, at time instance $t=1$, we draw $\bx_{1}\hspace{-0pt}\in\hspace{-0pt}\mathbb{R}^{n}$ from
the standard normal distribution with $s_{0}$ nonzero elements. Then, we generate a sequence of correlated sparse vectors $\bx_{t}$, $t=\{2,3,\dots,d+q\}$, where each $\bx_{t}$ satisfies $\|\bx_{t}-\bx_{t-1}\|_{0}=s_{0}/2$, where $\|\cdot\|_{0}$ denotes the number of
nonzero elements of a given vector. As this could lead to $\|\bx_{t}\|_{0}> s_{0}$, we add the constraint $\|\bx_{t}\|_{0}\in[s_{0},s_{0}\hspace{-0pt}+\hspace{-0pt}15]$. Whenever $\|\bx_{t}\|_{0}\hspace{-0pt}>\hspace{-0pt}s_{0}\hspace{-0pt}+\hspace{-0pt}15$, we reset $\bx_{t}$ to $\|\bx_{t}\|_{0}\hspace{-0pt}=\hspace{-0pt}s_{0}$ by setting $\|\bx_{t}\|_{0}-s_{0}$ randomly selected positions to zero. 
Thirdly, we initialize the prior information; in order to address real scenarios,
where we do not know the sparse and low-rank components, we use the batch-based
RPCA \cite{CandesRPCA} method to separate the training set $\bM_{0}=[\bx_{1}+\bv_{1}~...~\bx_{d}+\bv_{d}]$
so as to obtain $\bB_{0}=[\bv_{1}~...~\bv_{d}]$. In this experiment, we use
three (a.k.a., $J=3$) sparse components as prior information and we set $\bZ_{0}:=\{\mathbf{0},\mathbf{0},\mathbf{0}\}$.
%

We then evaluate the CODA method on the test set of vectors $\bM=[\bx_{d+1}+\bv_{d+1}~...~\bx_{d+q}+\bv_{d+q}]$. We vary 
$s_{0}$ (from 10 to 110) and the number of measurements $m$, and we assess the probability of success for the sparse $\mathrm{Pr_{\text{sparse}}(success)}$ and the low-rank~$\mathrm{Pr_{\text{low-rank}}(success)}$ component, averaged over the test
vectors. $\mathrm{Pr_{\text{sparse}}(success)}$ (resp. $\mathrm{Pr_{\text{low-rank}}(success)}$) is defined as the number of times in which the sparse component $\bx_{t}$ (resp. the low-rank component~$\bv_t$) is recovered within an error $\|\bhx_{t}-\bx_{t}\|_{2}/\|\bx_{t}\|_{2}\leq10^{-2}$ (resp. $\|\bhv_{t}-\bv_{t}\|_{2}/\|\bv_{t}\|_{2}\leq 10^{-2}$)
divided by the total 50 Monte Carlo simulations. \textcolor{black}{In Algorithm~\ref{CODAAlg}, we have set $\epsilon=0.8$, $\lambda=1/\sqrt{n}$,~$\mu=10^{-3}$, and the number of clusters $C=7$.} 

The results in Fig. \ref{corpcaPerform} demonstrate the efficiency of the proposed CODA. In Fig. \ref{corpcaPerform}, $\mathrm{Pr(success)}$ is measured and visualized in the bone color that the scale [see Fig. \ref{scale}] is proportional to $\mathrm{Pr(success)[\%]}$, i.e., from black ($0\%$ success) to white ($100\%$ success). CODA can recover the 500-dimensional data from small measurements rates [$m/n = 0.25$ to 0.6, see the white areas in Fig. \ref{codaPerformBg}]. For values of $s_0>70$, CORPCA-$n$-$\ell_{1}$ can not recover the sparse components successfully [see grey areas in Fig. \ref{corpcaPerformBg}], while CODA-$n$-$\ell_{1}$ succeeds. Fig. \ref{reprocsPerform} shows that the performance of ReProCS is worse than that of CODA-$n$-$\ell_1$. Moreover, Fig. \ref{grastaPerform} shows that GRASTA delivers lower low-rank recovery performance than CODA-$n$-$\ell_1$. 

\vspace{-0.14pt}
\subsection{Compressive Video Foreground-Background Separation}
\label{sec"videoResults}
\vspace{-0.4pt}
\begin{figure*}[t!]
	\centering
	\subfigure[CODA: \texttt{Bootstrap}]{
		\includegraphics[width=0.45\textwidth]{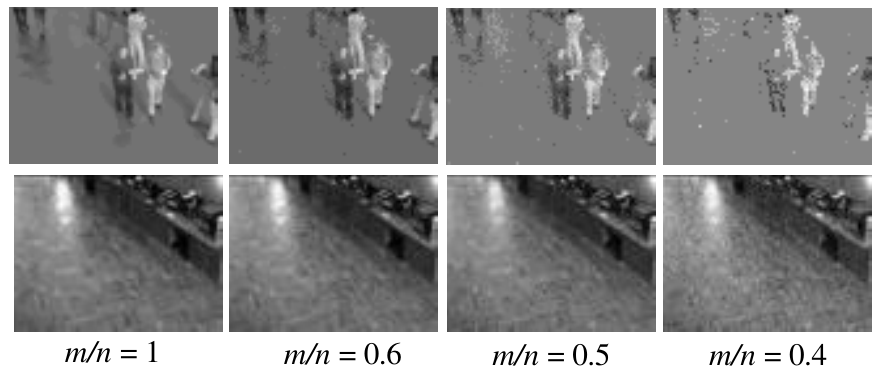}\label{bootstrapCompressedCoda}}
	\vspace{-0.8pt}
	\subfigure[CODA: \texttt{Curtain}]{
		\includegraphics[width=0.42\textwidth]{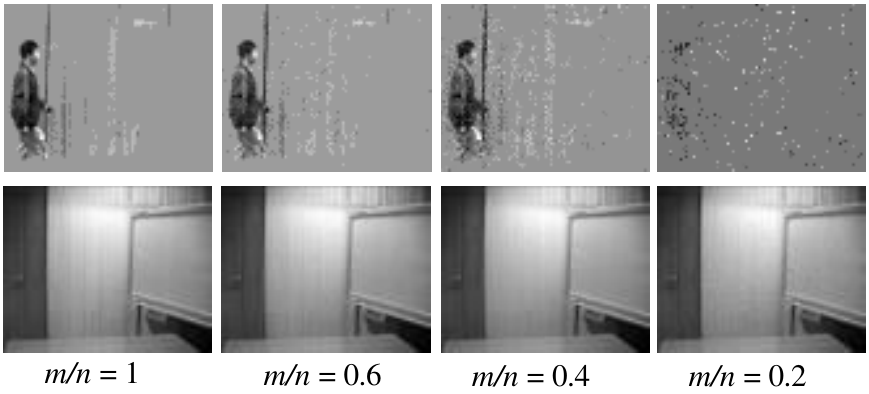}\label{curtainCompressedCoda}}
	\vspace{-0.8pt}
	\subfigure[CORPCA \cite{LuongGlobalSIP17}: \texttt{Bootstrap}]{
		\includegraphics[width=0.45\textwidth]{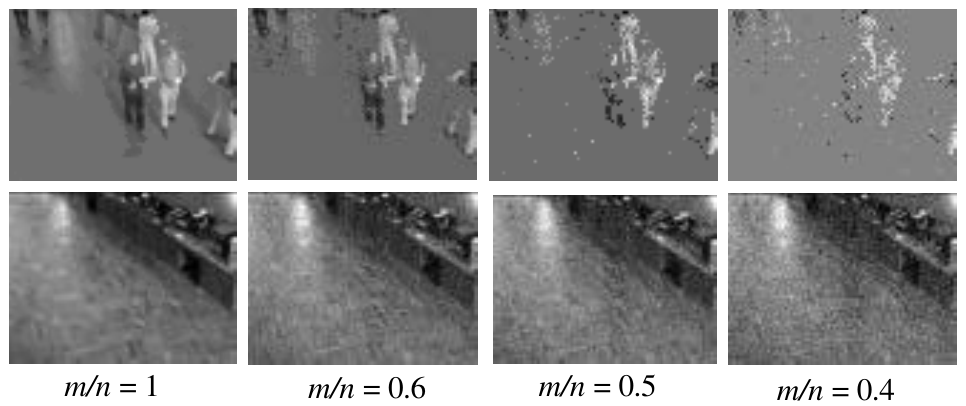}\label{bootstrapCompressedCorpca}}
	\subfigure[CORPCA \cite{LuongGlobalSIP17}: \texttt{Curtain}]{
		\includegraphics[width=0.43\textwidth]{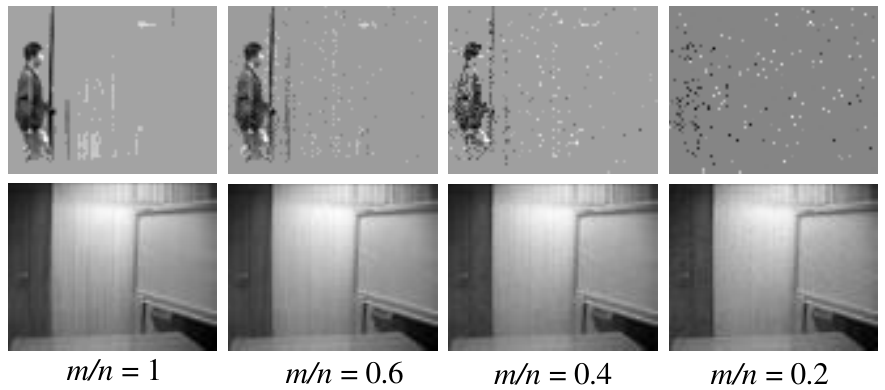}\label{curtainCompressedCorpca}}
	\vspace{-0.2pt}
	\subfigure[ReProCS \cite{GuoQV14}: \texttt{Bootstrap}]{
		\includegraphics[width=0.45\textwidth]{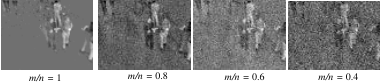}\label{bootstrapCompressedReprocs}}
	\subfigure[ReProCS \cite{GuoQV14}: \texttt{Curtain}]{
		\includegraphics[width=0.43\textwidth]{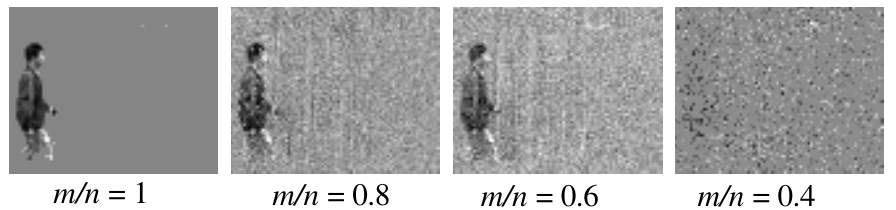}\label{curtainCompressedReprocs}}
	\caption{Foreground recovered using CODA, CORPCA, and ReProCS under different measurement
		rates: (a) \texttt{Bootstrap} (frame no. 2213) and (b) \texttt{Curtain} (frame
		no. 1766).
		\vspace{-0.0pt}
	}\label{figVisualCompressed}
\end{figure*}

We now assess our CODA method in the application of compressive video background-foreground separation using real video content and compare it against CORPCA \cite{LuongGlobalSIP17}, GRASTA \cite{JHe12}, and ReProCS \cite{GuoQV14}. We consider two video sequences~\cite{Li04}, \texttt{Bootstrap} (rescaled to 60$\times$80 pixels) and \texttt{Curtain} (rescaled to 64$\times$80 pixels), and use the first $d=100$ frames for training and the subsequent frames for evaluation. We use three sparse components as prior information: $\hat{\bx}_{t-1}$, $\hat{\bx}_{t-2}$, and $\hat{\bx}_{t-3}$. We consider the immediately previously reconstructed foreground as the first side information signal, i.e., $\bz_1=\hat{\bx}_{t-1}$. The other two side information signals are formed by applying motion-compensated extrapolation using the three previously reconstructed frames. Specifically, we perform forward optical-flow-based~\cite{LuongSoICT17,TBrox11} motion estimation from $\hat{\bx}_{t-2}$ to $\hat{\bx}_{t-1}$ (resp., $\hat{\bx}_{t-3}$ to $\hat{\bx}_{t-1}$) and then apply the motion vectors on $\hat\bx_{t-1}$ to generate $\bz_2$ (resp. $\bz_3$).


Fig.~\ref{figVisualCompressed} presents the results of CODA under various compressive rates $m/n$. The results show that we can recover the foreground and background more structural even by accessing a small number of measurements; for instance, with $m/n=0.5$ and $m/n=0.4$ for \texttt{Bootstrap} [Fig. \ref{bootstrapCompressedCoda}] and \texttt{Curtain} [Fig.~\ref{curtainCompressedCoda}], respectively. \texttt{Bootstrap} requires more measurements than \texttt{Curtain} due to the more complex foreground. It is clear that the results of CORPCA \cite{LuongGlobalSIP17} are worse than those of CODA (Figs. \ref{bootstrapCompressedCorpca}, \ref{curtainCompressedCorpca}) and the foreground images recovered with ReProCS \cite{GuoQV14} have a poor quality compared to CODA, even at a high rate $m/n=0.8$ (Figs. \ref{bootstrapCompressedReprocs}, \ref{curtainCompressedReprocs}). 
\vspace{-0.14pt}
\section{Conclusion}\label{conclusion}
\vspace{-0.9pt}
This paper proposed a compressive online decomposition algorithm (CODA) employing an $n$-$\ell_{1}$ cluster-based minimization that decomposes streaming data from compressive measurements. CODA incorporates multiple prior information in the decomposition problem and leverages the sparse structures via iteratively clustering and re-weighting the sparse components during the minimization. Numerical and compressive video foreground-background separation results have shown the efficiency of CODA compared to the existing methods. 

\vspace{-0.19pt}
\vspace{-0.8pt}
\bibliographystyle{IEEEtran}
\small
\bibliography{IEEEfull,IEEEabrv,bibliography}

\end{document}